# Single-spin readout for buried dopant semiconductor qubits

L.C.L. Hollenberg, C.J. Wellard, C.I. Pakes, and A.G. Fowler

*Centre for Quantum Computer Technology*
*School of Physics, University of Melbourne, VIC 3010, Australia*

## Abstract

In the design of quantum computer architectures that take advantage of the long coherence times of dopant nuclear and electron spins in the solid-state, single-spin detection for readout remains a crucial unsolved problem. Schemes based on adiabatically induced spin-dependent electron tunnelling between individual donor atoms, detected using a single electron transistor (SET) as an ultra-sensitive electrometer, are thought to be problematic because of the low ionisaton energy of the final $D^-$ state. In this paper we analyse the adiabatic scheme in detail. We find that despite significant stabilization due to the presence of the $D^+$, the field strengths required for the transition lead to a shortened dwell-time placing severe constraints on the SET measurement time. We therefore investigate a new method based on resonant electron transfer, which operates with much reduced field strengths. Various issues in the implementation of this method are also discussed.

Solid-state qubit systems where the quantum information is encoded by single donor degrees of freedom have been the subject of intense activity in recent years since the seminal work of Kane[1]. Indeed, recent advances in single-atom doping of semiconductors[2] bring closer to reality the construction of such a quantum computer (QC) device based on single donor spin[1,3] or charge[4] qubits. Solid-state qubits are of wide interest because of the nexus to scaleable fabrication technology, and in the case of spin based architectures, the relatively long coherence times of donor nuclear and electron spins. Progress towards the fabrication of such devices is reviewed in Ref 5. Detection of a single donor electron or nuclear spin remains a crucial problem, and a number of schemes have been proposed[1,6-8]. Recent direct measurements by Mamin et al[9] of few electron spins in $SiO_2$ are promising, with claims that the technique may be extendable to the single electron spin level. An interesting new scheme by Friesen et al[10] based on resonant spin to orbital conversion for electron spins confined in asymmetric quantum dots might also be applicable to the buried dopant case with appropriate gating. In this paper we report on investigations of the dynamics of the original indirect single spin detection proposal[1] for buried dopant systems, and as a direct consequence we develop a resonant transfer scheme suitable for both electron and nuclear spin readout.

Indirect spin detection involves transfer of the spin information to a spin-dependent charge transfer tunnelling process, which is detected using a single electron transistor (SET) as an ultra-sensitive electrometer[11]. A realistic architecture for spin-dependent charge transfer, applicable for electron and nuclear spin readout, is shown in Figure 1 for the case of phosphorous dopants in silicon. The concept relies on the adiabatic application of a DC electric field $F_0$ to induce tunnelling of the qubit electron to a secondary (spin polarised) "SET-donor", forming a doubly occupied $D^-$ donor state. Detection of the charge distribution change by the SET is effectively a measurement of the qubit spin state as the tunnelling event $D^0D^0 \rightarrow D^+D^-$ will be Pauli blocked if the qubit and SET-donor electron spins are parallel.

A potential problem with this scheme is the possible destruction of the qubit through ionisation due to shallow nature of the $D^-$ state (1.7meV binding energy). At typical operational temperatures (100mK), one may neglect Poole-Frenkel thermal ionisation, however, the question arises as to whether the $D^-$ state will tunnel to the conduction band under the application of the DC field destroying the qubit system during the SET measurement. For the small charge levels to be measured ($\Delta q < 0.05e$) single-shot RF SET readout operating near the quantum limit can obtain measurement times of order $T_{SET} \approx 1\mu s$[12], and determines the required survival time of the doubly occupied state. Given the experimental and theoretical complexity of these controlled single and double dopant systems, this crucial question of qubit survival over SET measurement timescales for the adiabatic indirect detection method has remained open.

In this paper we address this problem by carrying out detailed simulations of the device shown in Figure 1. Our starting point is a simple WKB calculation of the critical DC field strength, $F_0^*$, corresponding to a "safe" $D^-$ state dwell time of $T_{D^-} \approx 10\mu s > T_{SET}$. This serves as the main constraint for the spin dependent transfer scheme. We then compute the DC field strength $F_0^{ad}$ required for adiabatic spin transfer (for donor spacings 20 and 30nm) for direct comparison. Despite stabilization effects from the electron-hole interaction in the $D^+D^-$ system, which can be quite significant, we find $F_0^{ad}$ exceeds $F_0^*$ for all cases by more than an order of magnitude indicating that, even allowing for the crude approximation used in the determination of the critical field strength, the qubit will probably not survive adiabatic spin readout.

Using the framework developed for simulating the device in Figure 1, we investigate a resonant transfer readout scheme, similar to that implied by Laroniov et al[13] in the context of a $D^-$ based quantum computer proposal. In this work, however, we simulate the device shown in Figure 1 with a rapidly varying electric field applied to the gates. In this analysis the local DC fields required for addressing individual qubits for readout are shown to be much less than $F_0^*$. There is real potential for this approach in that resonant transfer may be achieved with state-of-the-art FIR laser technology[14] – a claim supported by the fact that the transition $D^0D^0 \rightarrow D^+D^-$ has been experimentally observed in spectroscopic studies[15] of bulk-doped Si:P. Finally, we analyse the resonant readout scheme for the case of nuclear spin readout involving a pre-readout preparation stage where the nuclear spin information is transferred to the electron spin.

We first consider the case of an isolated $D^-$ system. When a DC field $F_0$ is applied to the shallow $D^-$ one of the electrons can tunnel to the conduction band through the relatively low barrier formed. The field tunnelling dwell time $T_{D^-}(F_0)$ for such states is notoriously sensitive to the field strength $F_0$, indicating the difficulty in calculating this quantity. To set a reference scale for this initial analysis use a straightforward estimate for tunnelling through a Coulomb potential in the standard 1D WKB approximation[16]. At 100mK we solve the condition $T_{D^-}(F_0^*) = 10\mu s$ in the device configuration shown and obtain $F_0^*[D^-] = 0.00037 \text{Ry}/a_B$. In itself this is a small value, and would seem at first sight to be very discouraging. However, in the case of the $D^+D^-$ system there will be significant stabilization due to the electron-hole interaction. In order to compute $F_0^*$ for the $D^+D^-$ system and compare with the field strength $F_0^{ad}$ for adiabatic transfer we must solve the non-trivial donor molecular problem[17] in the presence of an electric field for the process $D^0D^0 \rightarrow D^+D^-$.

For the typical donor separations we will be considering ($R > 20$nm) the natural choice for a basis of states in the singlet sector are $\{|\psi_i\rangle\} = \{|LL\rangle, |LR\rangle, |RR\rangle\}$ (where $L$ and $R$ refer to the position of the electrons on left and right donors respectively). The system is initially in the $|LR\rangle = |D^0D^0\rangle$ state. A first order quantitative analysis of the adiabatic readout scheme is most easily carried out in the envelope approximation. Further improvements using Bloch structure of the donor electron wave functions are possible, however to illustrate the basic principles, this approach is quite adequate. We use the following wave functions to describe this basis of states:

$$f_{LL}(r_1, r_2) = N_{LL}(e^{-\alpha r_1}e^{-\beta r_2} + e^{-\beta r_1}e^{-\alpha r_2})(1 + \lambda r_{12})$$
$$f_{LR}(r_1, r_2) = N_{LR}(e^{-r_1}e^{-|r_2 - R|} + e^{-|r_1 - R|}e^{-r_2})$$
$$f_{RR}(r_1, r_2) = N_{RR}(e^{-\alpha|r_1 - R|}e^{-\beta|r_2 - R|} + e^{-\beta|r_1 - R|}e^{-\alpha|r_2 - R|})(1 + \lambda r_{12})$$

where the $N_{ij}$ are normalisation constants. For the case of the doubly occupied states we have taken the Chandrashakar wave function – well known to give a good account of the $H^-$ ion – where $\alpha$, $\beta$ and $\lambda$ are variational parameters ($\lambda$ controls the electron correlation strength). The wave functions are evaluated with standard parameters appropriate for donors in Si, i.e. $a_B = 2$nm, and $m_{eff} = 0.2 m_e$ (the scheme can be readily reworked for other substrate:dopant systems), and we scale the Rydberg energy to the $D^0$ ground state: $1\text{Ry} = 45.5\text{meV}$.

The bare (i.e. ungated) two-donor Hamiltonian matrix in the $\{|LL\rangle, |LR\rangle, |RR\rangle\}$ basis is formed by computing overlap integrals for various donor separations using Monte-Carlo quadrature. For large donor separations the basis of states is approximately orthogonal and the eigenstates of the system are:

$$|f_1\rangle \approx |LR\rangle, |f_2\rangle \approx \frac{1}{\sqrt{2}}[|LL\rangle - |RR\rangle], |f_3\rangle \approx \frac{1}{\sqrt{2}}[|LL\rangle + |RR\rangle]$$

To the bare two-donor Hamiltonian we also add a time dependent electric field along the inter-donor axis (defined to be the x-axis) of the form: $F(t) = F_0 + F_1 \sin \omega t$.

Considering adiabatic transfer first we take the DC case and set $F(t) = F_0$. The energy level diagram as a function of DC field strength $F_0$ is shown in Figure 2, for donor separations of $R = 20$nm and 30nm. As the donor separation is increased several observations can be made: the level crossing point $F_0^c(R)$ between the ground and excited states moves to lower values of $F_0$, the gap at the crossing narrows, and the binding energy of the $D^+D^-$ state at zero field decreases. For $F_0 < F_0^c$, the energy gap $DE = E_2 - E_1$ is of course a decreasing function of $F_0$ and has a residual $R$ dependence in its slope.

The single electron binding energy is computed by subtracting $E_{D+D^-}$ from the single electron state energy $E_{D+D0}$, giving the effective barrier height for the WKB tunnelling analysis. The results for the various characteristic field strengths are summarised in Table 1. As a measure of the adiabatic field required we define $F_0^{ad}$ as the field strength at which transition $|LR\rangle \to |LL\rangle$ has occurred with Prob$[|LL\rangle] > 0.99$ in the final (ground) state. The results, albeit for a different parameter set, agree qualitatively with the calculations of Fang et al[17] for the adiabatic transition. Clearly, in comparison to the isolated case $F_0^*[D^-] = 0.00037$Ry/$a_B$, the electron-hole interaction stabilizes the $D^+D^-$ state to a large extent against field-dependent tunnelling (stabilisation of the $D^+D^0$ state was also taken into account). Near the avoided level crossing, $F_0 > 0.040$Ry/$a_B$, the $D^+D^-$ state dwell time decreases to less than $10^{-6}$µs. Although larger separations seem to be favoured, the requirement $F_0^{ad} < F_0^*$ for qubit survival of the adiabatic transfer process is clearly not satisfied.

Of course, these conclusions depend on not only the veracity of the WKB calculation of $F_0^*$, which will receive corrections from transverse momenta[18] (tending to decrease the tunnelling probability in the forward direction), but also the extent to which electron-hole stabilization leads to $F_0^*[D^+D^-] \gg F_0^*[D^-]$. This statement is dependent on the complexities of the molecular description of the two donor system. However, given the order of magnitude difference in $F_0^*[D^+D^-]$ and $F_0^{ad}$ it would seem unlikely that improvements in the calculations will change the overall picture that adiabatic single spin readout, at the very least, severely tests the realm of current SET technology.

We thus turn our attention to the case of spin detection through resonant electron transfer, which as we will see can be carried out using much lower field strengths. To this end, we switch on the AC component $F_1$ with a small DC offset $F_0 < F_0^*$. If the oscillating field is set resonant with the gap $DE(F_0)$ Rabi oscillations $|LR\rangle \leftrightarrow |LL\rangle$ occur over a timescale controlled by the field amplitude $F_1$. After the Rabi time $T_{Rabi}$ the $|LR\rangle$ component goes to zero and the AC component of the field is switched off leaving the DC component on. Since the offset DC field merely allows for qubit selection and symmetry breaking, the value of $F_0$ can be quite small giving rise to a long enough $D^-$ state dwell time for SET measurement to take place.

The parameter regime for the process $D^0D^0 \to D^+D^-$ is shown in Figure 3. The energy gap for the process has been calculated as a function of DC field strength $F_0$ for two donor separations and plotted with respect to the single donor levels. The 1s-$2p_0$ and 1s-$2p_\pm$ transitions with non-zero dipole matrix element (solid horizontal lines) serve as natural boundaries for the resonant charge transfer process. For $R = 20$nm the transition is mainly below the 1s-$2p_0$ line, while for $R = 30$nm it occurs almost entirely in the $2p_0$-$2p_\pm$ gap.

In Figure 4 we show the results of a time dependent calculation, in this case an example of the transition $|LR\rangle \to |LL\rangle$ for $R = 30$nm occurring in the $2p_0$-$2p_\pm$ band. The local DC field was kept well below $F_0^*$ at $F_0 = 0.001$Ry/$a_B$. At $t = 20$ps the AC component was pulsed to $F_1 = 0.002$Ry/$a_B$ and held at that value over $T_{Rabi} = 98$ps during which the transition $|LR\rangle \to |LL\rangle$ takes place. Similar results were found for $R = 20$nm, although in avoiding the single donor levels slightly higher DC field strengths are required.

Single dopant implantation technology has a reached the stage where the device shown in Figure 1 can be fabricated[5] with a twin RF SET capability allowing for signal correlation and noise rejection[12]. However, the AC-gated version of readout based on resonant transfer is probably not experimentally feasible as it requires voltage pulse timing at frequencies ≈10THz. However, this analysis does contain all the essential ingredients for an optically based resonant transfer scheme where individual qubits are brought into resonance with a field tuned to the gap $DE(F_0,R)$ using the local DC gates, and thus selectively read out. Advances in FIR technology are bringing this wavelength regime into reach[14]. Observations of the $D^0D^0 \rightarrow D^+D^-$ transition for bulk doped Si:P ($1.7 \times 10^{17}$ cm$^{-3}$) shows a broad IR absorption peak at about 30meV[15], which agrees with our results for $F_0 = 0$ given that the mean donor separation for that dopant density is about 10nm. Photo-ionisation, over a timescale $T_{Photo}$ can be neglected by selecting $F_1$ to ensure $T_{Photo} >> T_{Rabi}$.

Finally, we consider the preparation stage for the case of nuclear spin readout, wherein the transfer of nuclear spin information to electrons takes place using the exchange interaction controlled by the J-gate. The original Kane proposal calls for the ability to adiabatically increase the exchange interaction between qubit and SET donors above $J_c = 0.058$ meV[19]. The two-donor spin system exhibits energy level crossings of a number of eigenstates at $J_c$. When $J$ is increased adiabatically through $J_c$ anti level-crossing behaviour maps the system eigenstates $|e_1e_2\rangle \otimes |n_1n_2\rangle$ (assuming spin polarized electrons in a $B = 2T$ field) as follows: $|\downarrow\downarrow\rangle|11\rangle \rightarrow |\downarrow\downarrow\rangle|11\rangle$, $|\downarrow\downarrow\rangle|10\rangle \rightarrow |\downarrow\downarrow\rangle|s_n\rangle$, $|\downarrow\downarrow\rangle|01\rangle \rightarrow |a_e\rangle|11\rangle$, $|\downarrow\downarrow\rangle|00\rangle \rightarrow |a_e\rangle|a_n\rangle$. Here $|s\rangle$ and $|a\rangle$ refer to symmetric and anti-symmetric states respectively. Note that if $|n_l\rangle$ is in the $|1\rangle \equiv |\downarrow\rangle$ state the electrons remain spin down whereas if $|n_l\rangle = |0\rangle$ the two-electron state is mapped to $|a_e\rangle$. The final $|e_1e_2\rangle$ state is thus dependent on the state of $|n_l\rangle$ and readout of this spin proceeds according to the spin dependent tunnelling scheme.

In order to determine the feasibility and fidelity of the preparation process, time dependent simulations of this process were carried out with the dephasing time of the nuclei ($t_n$) and electrons ($t_e$) included in a similar manner to previous simulations of the nuclear spin CNOT gate[20]. Assuming $J$ can be varied from 0.054meV to 0.063meV over 12μs, Figure 5 shows the fidelity of the readout preparation operation as a function of $t_n$ and $t_e$ dephasing times. Given that $t_n >> t_e$, the electron dephasing dominates the fidelity of the process. Recent measurements[21] in isotopically pure Si$^{28}$ give $t_e > 60$ms at 7K.

A calculation of the exchange coupling $J$ using Kohn-Luttinger type wave functions deformed by a surface bias applied to a J-gate was carried out in Ref 22. For a bias of 1.0V applied to a surface gate above donors in the geometry of Figure 1 the exchange coupling was found to be 0.030meV for $R = 20$nm. At face value this would be insufficient to access the desired level crossing, however, the voltage dependent exchange coupling calculation is the subject of ongoing work by a number of groups, and we would not claim this to be the last word. We note here also that the relevant spin information could be transferred using nuclear spin dependent Rabi flipping of the donor electron. Let $E_{Z\uparrow}$ ($E_{Z\downarrow}$) denote the Zeeman energy of the donor electron when the nuclear spin is up (down). The magnetic field frequency resonant with the $E_{Z\downarrow}$ transition and the maximum field strength such that $E_{Z\uparrow}$ is off resonance can be calculated from $E_{Z\downarrow} \cong 0.116$meV and $E_{Z\uparrow} - E_{Z\downarrow} = 4A$ (where $A \cong 1.2 \times 10^{-4}$ meV is the nucleus-electron hyperfine interaction energy). For an RF field of magnitude $|B_{ac}| \cong 10^{-4}$ T we obtain a relatively fast electron flipping time of 0.17μs.

In summary, we have analysed the adiabatic single spin readout scheme and found that within the approximations employed the condition for qubit survival $F_0^{ad} < F_0^*$ – i.e that the applied field is less than the maximum critical field – is not satisfied for typical donor separations 20 – 30nm. However, the resonant scheme was found to be a viable alternative involving DC fields much less than the critical field.

This work was supported by the Australian Research Council, and the Victorian Partnership for Advanced Computing. The authors would like to acknowledge discussions with R. Scholten, M. Testolin, A. Hamilton, A. Greentree and J. McCallum.

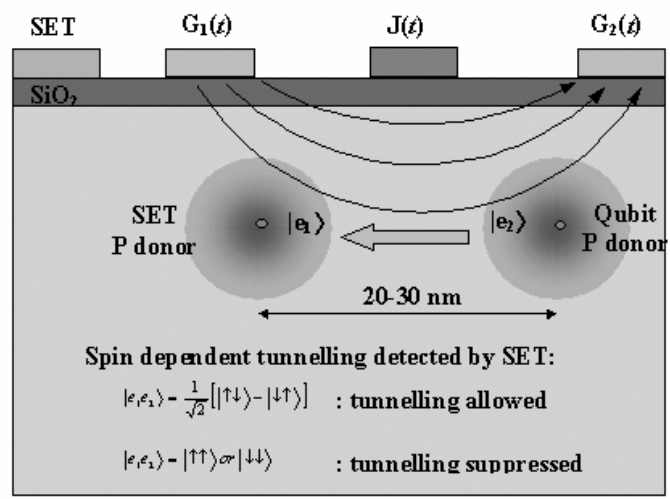

**Figure 1** Spin dependent charge transfer scheme for single-spin readout (electron or nuclear spin) based on single dopant atoms in semiconductors.

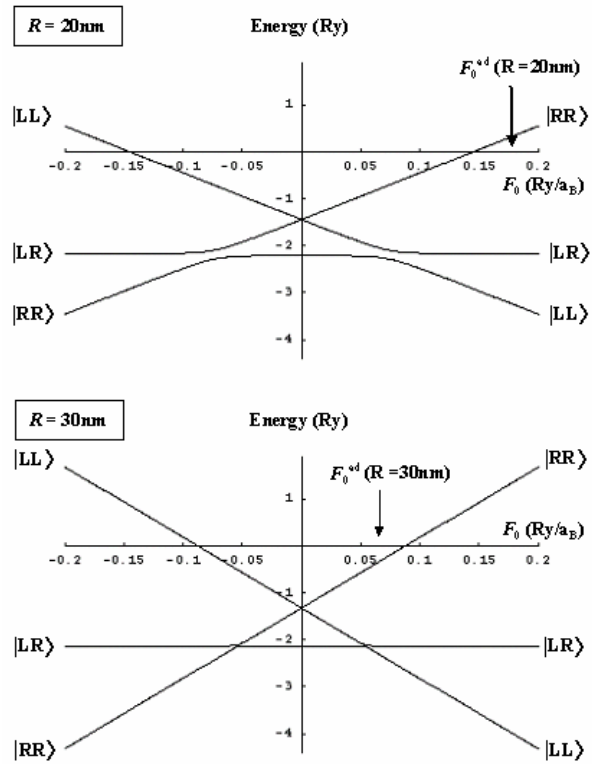

**Figure 2** Energy level diagrams for $R = 20$nm and 30nm as a function of the DC field strength (avoided level crossings at zero field are not resolved on these plots).

| $R$ (nm) | $F_0^c$ | $F_0^{ad}$ | $F_0^*$ |
|---|---|---|---|
| 20 | 0.0737 | 0.1740 | 0.0087 |
| 30 | 0.0540 | 0.0630 | 0.0058 |

**Table 1** DC field strengths corresponding to the level crossing ($F_0^c$), adiabatic transition $D^0D^0 \rightarrow D^+D^-$ ($F_0^{ad}$), and 10μs $D^+D^-$ dwell-time ($F_0^*$). All values are quoted in Ry/$a_B$.

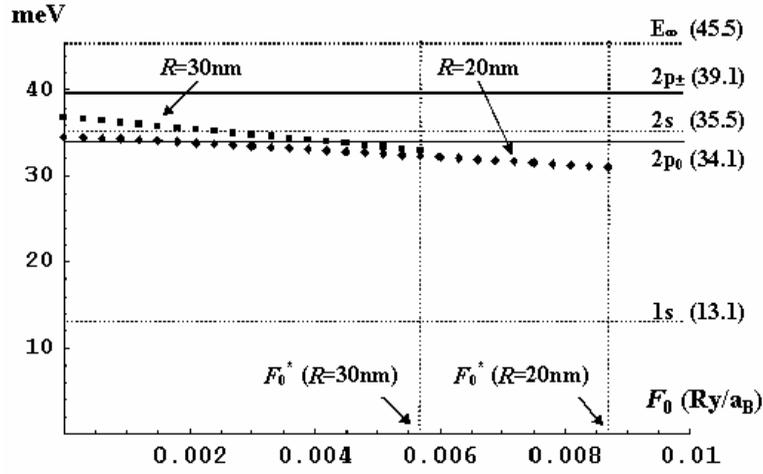

**Figure 3** Energy level diagram (relative to the donor ground state) showing the $D^0D^0 \to D^+D^-$ ($|LR\rangle \to |LL\rangle$) transition energy as a function of DC field $F_0$ for $R = 20$nm (diamonds) and $R = 30$nm (squares) in comparison to the low-lying single donor levels. The relevant 1s-2p transitions are shown as solid horizontal lines.

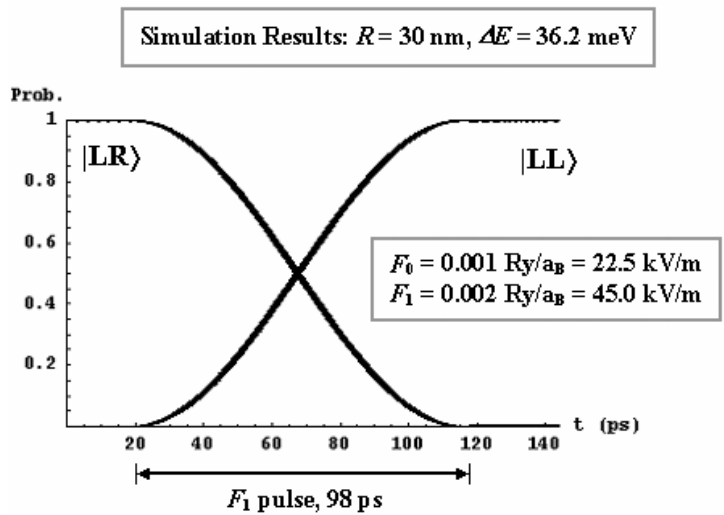

**Figure 4** Simulation results for the time dependent state probabilities during resonant transfer between donors.

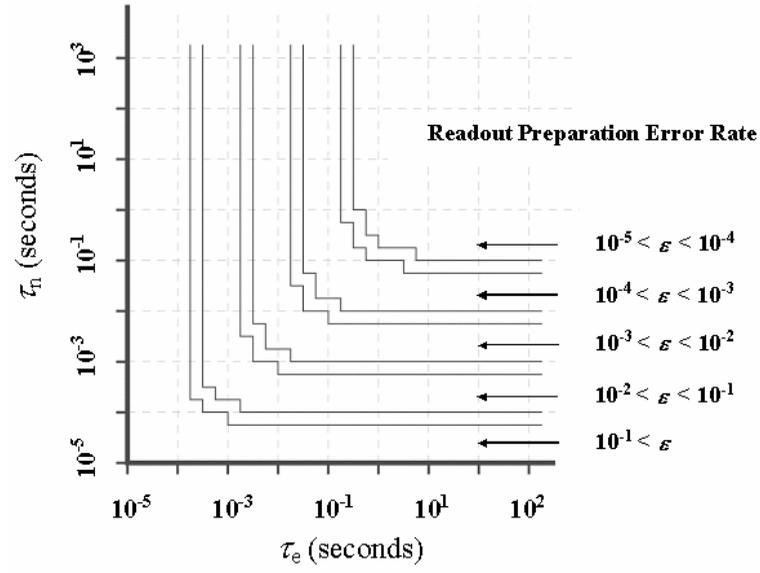

**Figure 5** Fidelity of adiabatically transferring nuclear spin information to the electrons of a two donor system for a range of dephasing times $t_n$, $t_e$.